\documentclass[12pt]{article}

\usepackage{graphicx}
\usepackage{url}
\usepackage[margin={1cm,-1cm},font=small,labelfont=bf,labelsep=endash]{caption}

\begin{document}

\title{Popularity, Novelty and Attention}
\author{Fang Wu and Bernardo A. Huberman\\HP Laboratories\\Palo Alto,
CA 94304}
\maketitle

\bigskip

\abstract{We analyze the role that popularity and novelty play in attracting the attention
of users to dynamic websites. We do so by determining the performance of three different strategies
that can be utilized to maximize attention. The first one prioritizes
novelty while the second emphasizes popularity. A third strategy looks
myopically into the future and prioritizes stories that are expected to generate
the most clicks within the next few minutes.  We show that the first two
strategies should be selected on the basis of the rate of novelty decay,
while the third strategy performs sub-optimally in most cases. We also demonstrate that
the relative performance of the first two strategies
as a function of the rate of novelty decay changes abruptly around
a critical value, resembling a phase transition in the physical world.}

\pagebreak

\section{Introduction}

As millions of people use the web for their social, informational,
and consumer needs, content providers vie for their limited
attention by resorting to a number of strategies aimed at
maximizing the number of clicks devoted to their web sites \cite{falkinger-07}. These
strategies range from data personalization and short videos to the
dynamic rearrangement of items in a given page, to name a few \cite{GKM-99,HTT-04,Zhang-00}. In
all these cases the ultimate goal is the same: to draw the
attention of the visitor to a website before she proceeds to the
next one \cite{HPPL-98}. Obviously, the more interesting and relevant the site
the more valuable it will be to users. In addition, since users
need to decide among the existing plethora of links and sites,
their popularities are a determinant of their success, for people often
click on given links for no other reason than the fact that many others
do. If we add the fact that without novelty attention tends to
decay in time, one has a first order list of the requirements for
capturing people's attention.

Within this context, we have recently shown that there is a strong
interplay between novelty and collective attention, which is
universally manifested in a rather swift initial growth of the
number of people looking at a new item within a site and its
eventual slowdown as interest fades among the population \cite{wu-huberman-07}.
This result suggests that ordering the links of a given page by
their novelty can guarantee a high degree of attention. This is
indeed the case in many news websites, notably \texttt{digg.com}.

And yet, given the role that popularity plays in attracting the
attention of users, a natural question arises as to whether
alternative orderings, like one giving priority to popularity over
novelty, might not do better at attracting viewers to a site.

This paper answers this question by taking the dynamics of
collective attention to a finer level of detail and examining the
role that popularity and novelty play in determining the number of
clicks within a given page. In particular, we study three different strategies
that can be deployed in order to maximize attention. The first strategy prioritizes
novelty while the second emphasizes popularity. The third strategy looks
myopically into the future and prioritizes stories that are expected to generate
the most clicks in the next few minutes.  We show that the first two
strategies should be selected on the basis of the rate of novelty decay,
while the third strategy performs sub-optimally in most cases. Most interestingly,
we discover that the relative performance of the first two benchmark strategies
as a function of the rate of novelty decay switches so sharply around
some critical value that it resembles phase transitions observed in
the real world.

The work is organized as follows. We first study the question of
whether or not the location of a link in a page determines the
overall number of clicks in a given time interval. Having answered
this in the affirmative through an empirical study of \texttt{digg.com}, we
then proceed to introduce a set of indexes whose values determine
the optimal strategy to be pursued in order to maximize attention
to a page. Using measured values of the rate of decay from
digg.com we built a realistic simulator to collect statistically
significant data to measure each of the indices
introduced.

We then study the performance of each of these indices as a
function of the decay rate and show which strategy optimizes
viewing for given values of the decay. Most importantly we compute
a full phase diagram that indicates at a glance the optimal
strategy to use given the parameter values of the site. This phase
diagram exhibits a sharp boundary between the choice of
prioritizing novelty over popularity, thus resembling a phase
transition.

Finally we summarize our results and discuss their implications for the design of dynamic websites.

\section{Location matters}

In this section we study how the order in which links are placed
within a webpage (e.g.~the news stories of \texttt{digg.com})
determines the number of clicks within a certain time frame.
Assume that time flows discretely as $t=0, 1, 2 \dots$ minutes.
Let $N_t$ denote the number of clicks, or \emph{digg number} of a story
in \texttt{digg.com}, that appeared on the website $t$ minutes ago (in this
case we say that the story has \emph{lifetime} $t$). As we showed
earlier \cite{wu-huberman-07} the growth of $N_t$ satisfies the following
stochastic equation:
\begin{equation}
\label{eq:benchmark-model}N_{t+1} = N_t (1+ a r_t X_t),
\end{equation}
where $r_t$ is a \emph{novelty factor} that decays with time and
satisfies $r_0=1$, $X_t$ is a random variable with mean 1, and $a$
is a positive constant.

This equation takes into account two important factors that
together determine the growth of collective attention: \emph{popularity} and \emph{novelty}. The popularity effect is captured by the multiplicative form of
Eq.~(\ref{eq:benchmark-model}), and the novelty effect is
described by $r_t$. All other factors are contained in the noise
term $X_t$.

We next take the analysis to a finer level by considering a third
\emph{position factor}. A news story displayed at a top position
on the front page easily draws more attention than a similar story
placed on later pages. Hence the growth decay $a r_t$ should
depend on the physical position at which the story is
posted.

In the specific case of \texttt{digg.com}, its front page is divided
into 15 slots, being able to display 15 stories at a time. The
stories are always sorted chronologically, with the latest story
at the top. If we label the positions from top to bottom by
$i=1,2, \dots, 15$, we can modify Eq.~(\ref{eq:benchmark-model}) to
allow for an explicit dependency of $a$ on $i$:
\begin{equation}
\label{eq:model}N_{t+1} = N_t (1+ a_i r_t X_t),
\end{equation}
where $a_i$ is a position factor that decreases with $i$.

The assumption that the novelty effect and the position effect can
be separated into two factors $r_t$ and $a_i$ needs to be tested
empirically. To this end we tracked the growth rate for each slot,
rather than for each story. For multiplicative models it is
convenient to define the logarithmic growth rate
\begin{equation}
\label{eq:log-growth-rate}s_t = \log N_{t+1} - \log N_t.
\end{equation}
When $a$ is small (which is always true for short time periods) we
have from Eq.~(\ref{eq:model})
\begin{equation}
s_t^i \approx a_i r_t X_t
\end{equation}
for a story placed at position $i$ at time $t$. Taking expectation
of both sides, we have
\begin{equation}
E s_t^i \approx a_i r_t,
\end{equation}
since $E X_t=1$.

\begin{figure}
\centering
\begin{minipage}{2.5in} \centering
\includegraphics[width=2.5in]{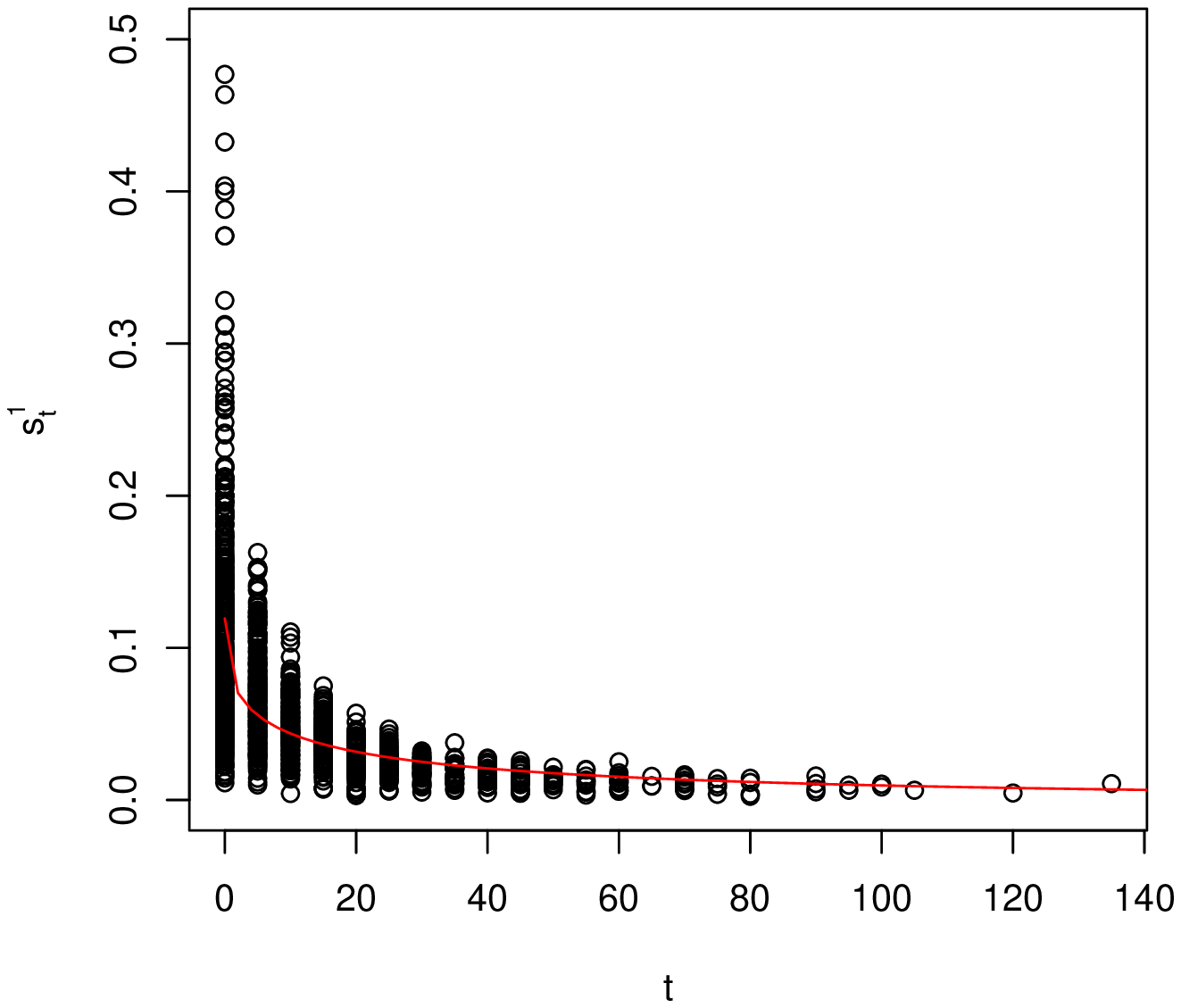}\\ {\small (a)}
\end{minipage}
\begin{minipage}{2.5in} \centering
\includegraphics[width=2.5in]{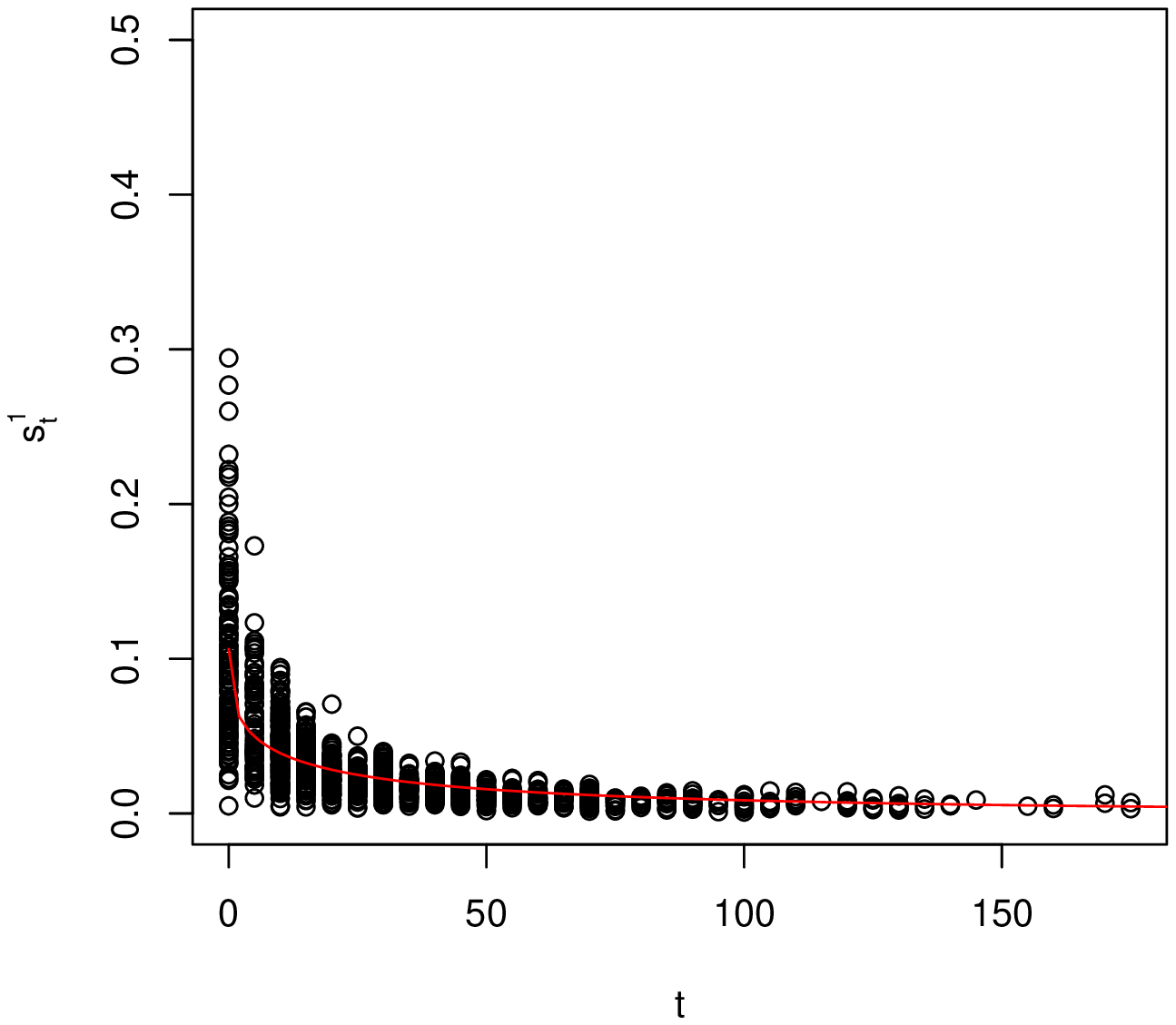}\\ {\small (b)}
\end{minipage}
\caption{\label{fig:st}The logarithmic growth rate for the top two
positions on the front page of \texttt{digg.com}. Time is measured
in minutes. Data is collected every 5 minutes, the rate at which the
front page is refreshed. The solid curve in (a) is the result of a
minimum mean square fit to the data (see text for more details). It
has the functional form $f(t)=0.120 \,e^{-0.4 t^{0.4}}$. The curve
in (b) has the functional form $f(t)=0.106 \,e^{-0.4 t^{0.4}}$.}
\end{figure}

\begin{figure}
\centering
\includegraphics[width=3in]{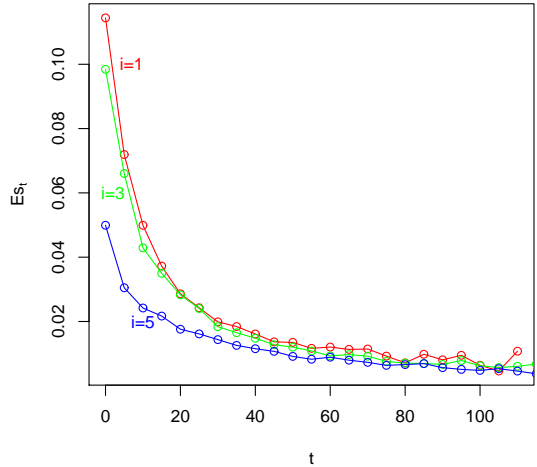}
\caption{\label{fig:s135}The expected logarithmic growth rate for
position 1, 3 and 5 on the front page of \texttt{digg.com}. Time is
measured in minutes. As can be seen, the growth rate decays as the
story moves to lower positions.}
\end{figure}

The logarithmic growth rate $s_t^i$ can be measured as follows.
For each fixed position $i$, if a \texttt{digg} story appears on
that position at both times $t$ and $t+5$ (the front page is
refreshed every 5 minutes), then the observed quantity $\frac 15
(\log N_{t+5} - \log N_t)$ counts as one sample point of $s^i_t$.
Fig.~\ref{fig:st}(a) plots 1,220 sample points collected from the
top position at various times. Fig.~\ref{fig:st}(b) is a similar
plot for the second top position. By comparing (a) and (b) we see
that $s^2_t$ indeed tends to fall below $s^1_t$, which indicates
that the position effect is real. To better illustrate the
position effect, we plot the expected growth rate for position 1,
3 and 5 in Fig.~\ref{fig:s135}. As can be seen there, the growth
rate decays as the story moves to lower positions.

From this data we can also determine the values of $a_i$
quantitatively. We already established that for digg.com the
precise functional form of the decay factor is $r_t=e^{-0.4
t^{0.4}}$. Thus, for these particular values, the minimum mean
square estimator $\hat a^i$ minimizes
\begin{equation}
\min_{a^i} \sum_j [s^i_{t_j}(j) - a^i r_{t_j}]^2 = \min_{a^i} \sum_j
[s^i_{t_j}(j) - a^i e^{-0.4 t_j^{0.4}}]^2,
\end{equation}
where $t_j$ is the lifetime of the $j$'th data point. The
estimator for the 1,220 data points obtained from the top position
is calculated to be $\hat a^1=0.120$. The fitted curve $\hat a^1
r_t = 0.120 e^{-0.4 t_j^{0.4}}$ is shown as a solid curve in
Fig.~\ref{fig:st}(a). An estimator $\hat a^2=0.106$ for the second
top position is also calculated and plotted in
Fig.~\ref{fig:st}(b). As can be seen from those figures, the
position effect ($a^i$) and the novelty effect ($r_t$) can indeed
be separated. We can then conclude that Eq.~(\ref{eq:model})
fits the data very well.

\section{Optimal ordering for maximal attention}

We now consider the order in which news stories should be
displayed on a web page so as to generate the largest number of
clicks within a certain time period $T$. This time period needs to
be finite because the total number of clicks diverges as $T$ goes
to infinity. Equivalently, in an infinite-horizon framework, we
could discount future clicks with a discount parameter $\delta$,
so that one click at time $t$ counts as $\delta^t$ click at time
0. The objective then is to maximize $\sum_{t=0}^\infty \delta^t
N_t$, where $N_t$ is the total number of clicks generated from the
news page in period $t$. In what follows we will consider the
finite-horizon objective.

To simplify the problem we confine ourselves to a subset of
ordering strategies called \emph{indexing strategies}, which
is defined as follows. Given a story's state, which in our model
is just a two-vector $(N_t, t)$, one first calculates an index $O$
for each story using a predefined \emph{index function} $O(N_t,
t)$, and then sorts the stories based on their indices. The story
with the largest index is displayed at the top, the story with the
second largest index next, and so on \cite{ninomora-02,wu-huberman-07a}.

Rather than considering a general index function we will
concentrate on three simple strategies. While neither of them is
perfect, each can increase overall attention to the site.

\begin{enumerate}
\item $O_1(t)=-t$. The stories are sorted by their novelty, with the
newest story at the top. This is what \texttt{digg.com} is doing
today.

\item $O_2(t)=N_t$. The stories are sorted by their popularity,
with the most popular story at the top. This strategy is based on
the fact that attention grows in a multiplicative fashion (popular
stories are more likely to become even more popular).

\item $O_3(t)=N_t r_t$. This is the ``one-step-greedy'' strategy.
Ignoring the position effect (assume $a=1$), a story in state
$(N_t, t)$ generates on average $N_t r_t$ more clicks (or ``diggs'' if
one considers \texttt{digg.com}) in the next period. This strategy thus
places the most ``replicated'' story at the top.
\end{enumerate}

\noindent Notice that because $N_t$ grows with time, the effect of sorting
by $O_1$ is almost the opposite of sorting according to $O_2$.

\begin{figure}
\centering
\includegraphics[width=3in]{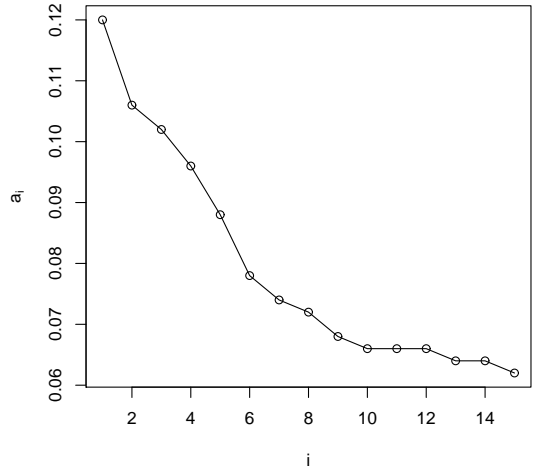}
\caption{\label{fig:ai}The position factor decays as the position
lowers. The values of $a_i$ are measured by tracking the 15 slots on
\texttt{digg.com}'s front page.}
\end{figure}

In order to test these strategies, we built a simulator that
closely resembles the functioning of \texttt{digg.com} in that it
incorporates the following rules:

\begin{enumerate}
\item Initially there are 15 stories, all in state
$(N_t, t)=(1,0)$. In words, each story starts with 1 digg and
lifetime 0. (Because our model is purely multiplicative, the initial
digg number does not matter. We just set it to be 1.)

\item \label{loop:start} Allocate the 15 stories to 15 positions,
in decreasing order of their $O(N_t, t)$, for any given index
function $O$.

\item Time evolves one step (5 minutes) at a time. The number of
diggs generated from a story at position $i$ is given by
\begin{equation}
\Delta N_{t+5} = N_{t+5}-N_t = 5 a_i r_t X_t N_t.
\end{equation}
The total number of diggs generated in this time step is the sum of
15 such numbers.

The values of $a_i$ were estimated from real data and shown in
Fig.~\ref{fig:ai}. $r_t=e^{-0.4 t^{0.4}}$. $X_t$ is randomly drawn
from a normal distribution with mean 1 and standard deviation 0.5
(obtained from the real data from \texttt{digg.com}).

\item On average every 20 minutes a new story arrives. Thus the
number of stories arriving in one time step (5 minutes) follows a
Poisson distribution with mean 0.25. When a new story enters the
pool, the story with the lowest index is dropped, maintaining 15
stories in total. (It is possible the a new story is dropped
immediately after its arrival if it happens to have the lowest
index.)

\item Go back to Step \ref{loop:start} until the loop has been
repeated for enough rounds.
\end{enumerate}

The performance of all three index functions were tested in our
simulator. For each index function, Steps 2 to 5 were repeated
100,000 times (or equivalently 500,000 minutes). Strategy $O_1$
(sort by novelty) achieved a total number of 514,314.8 diggs.
Strategy $O_2$ (sort by popularity) only generated 354.6 diggs.
Strategy $O_3$ (one-step-greedy) generated 452,402.3 diggs. Thus
for these parameter values $O_1$ turns out to be best strategy,
since it is $13.7\%$ better than $O_3$ and tremendously better
than $O_2$. This confirms that \texttt{digg.com} is using the
right strategy.

The reason for the poor performance of the index $O_2$ is easy to
understand. $O_2$ gives higher priority to stories that have been
dugg many times. According to the indexing rule, after one period
new stories can never find their way to the front page since all
the old stories have more than 1 digg! When novelty decays fast,
the old stories remaining on the front page soon lose their
freshness and cease to generate any new diggs. The system thus
gets frozen in an unfruitful state.

The fact that $O_1$ outperforms $O_3$ is a bit harder to
understand. Some intuition can be gained by considering an extreme
case. Suppose each story completely loses its novelty after one
second ($r_0=1$, $r_t=0$ for all $t>0$). Then only  ``new
arrivals'' should be displayed since they are the only ones that
can generate new diggs. Sorting stories by their lifetime is a
good idea when novelty decays fast. On the other hand, if novelty
never decays ($r_t \equiv 1$), the lifetime factor becomes
irrelevant. Thus in this case, strategy $O_3$, which prioritizes
popular stories, will win over $O_1$. Hence, the fact that $O_1$
works better than $O_3$ in our simulations shows that novelty
decays relatively fast for \texttt{digg.com}. Should it decay at a
slower rate, $O_3$ would be a better choice.

We point out that our simulation only showed that the ordering
implied by $O_1$ works better than $O_3$ for a particular choice
of $T$. In general this may not be true for other values of $T$.
In fact, for a time interval of $T=5$ minutes (one time step)
$O_3$ is by definition the best strategy. Hence, comparing the
performance of two or more index functions only makes sense after
one has specified a time horizon (or how much the future should be
discounted if an infinite horizon is assumed).

\begin{figure}
\centering
\includegraphics[width=5in]{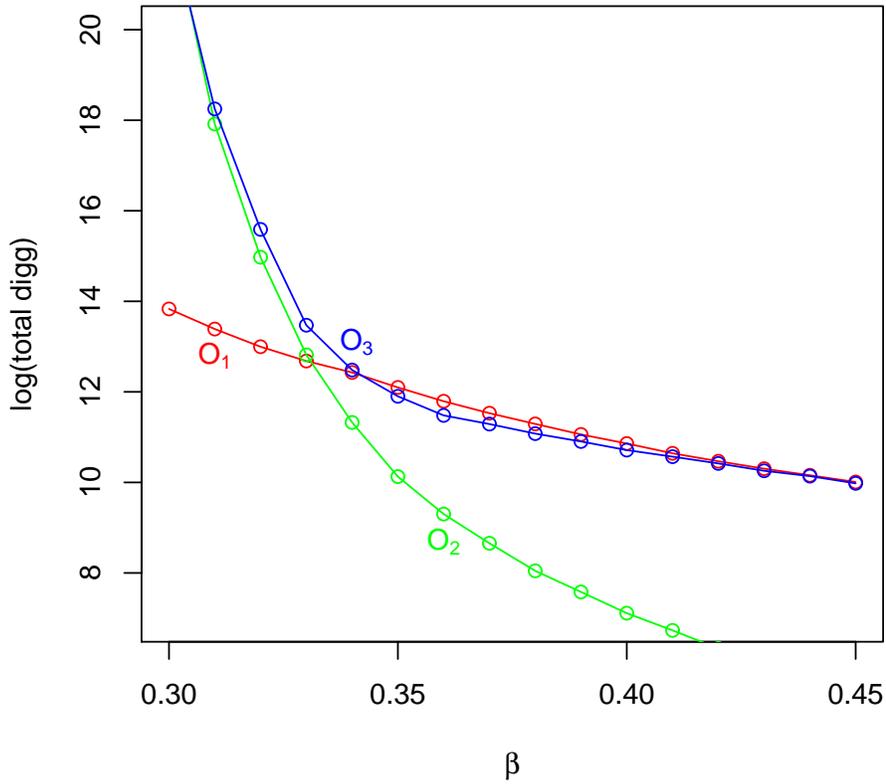}
\caption{\label{fig:beta}The total number of diggs generated using
three ordering strategies $O_1$, $O_2$, and $O_3$, for $\alpha=0.4$
and a range of $\beta$. The novelty factor decays as $r_t=e^{-\alpha
t^\beta}$. Performance is measured by the logarithm of the total
number of diggs generated in 10,000 time steps. As can be seen,
$O_3$ asymptotically approaches $O_1$ and $O_2$ in the fast decay
(large $\beta$) and slow decay (small $\beta$) limit, respectively.
A phase transition happens around $\beta=0.335$.}
\end{figure}

In order to quantitatively test the limiting behavior of the three
strategies, we repeated our simulations for a range of different
values of the decay parameter $r_t$. Our previous work suggested
that $r_t$ decays as a stretched exponential function, whose
general form can be written as $r_t= e^{-\alpha t^\beta}$. For
\texttt{digg.com} it turns out $\alpha=\beta=0.4$. The parameter
$\beta$ determines the decay rate. For fixed $\alpha$, the larger
$\beta$, the faster $r_t$ decays. We repeated our experiment for
$\alpha=0.4$ and $\beta \in [0.30, 0.45]$. The result is shown in
Fig.~\ref{fig:beta}. The performance of each indexing strategy is
measured by the logarithm of the total number of diggs generated
in 10,000 rounds. We see that as $\beta$ increases (faster decay),
the number of diggs decreases for all three indexing strategies.
When $\beta>0.34$, $O_1$ performs slightly better than $O_3$ and
much better than $O_2$. When $\beta<0.33$, however, $O_3$ and
$O_2$ perform significantly better than $O_1$. In other words, on
the two sides of the value of $\beta=0.335$, the stories should be
displayed in completely reversed order! We therefore say that a
\emph{phase transition} takes place at the value of $\beta=0.335$.

Other points worth mentioning are that in Fig.~\ref{fig:beta}
$O_3$ asymptotically approaches $O_1$ and $O_2$ both in the fast
and slow decay limits, and that in general $O_3$ is the best index
among the three strategies (although for the specific parameters
of \texttt{digg.com} ($\alpha=\beta=0.4$) and our particular time
horizon $O_1$ is slightly better). This is because $O_3$ trades
off between popularity and novelty instead of betting on only one
factor. To see this, consider the equivalent index function
\begin{equation}
O_3'(N_t,t) = \log O_3(N_t,t) = \log N_t + \log r_t.
\end{equation}
Clearly, $O_3'$ linearly trades off between $\log N_t$ and $\log
r_t$, assigning identical weight to the two effects. This is by no
means the best tradeoff. For example, the index function
\begin{equation}
O_4(N_t,t) = 0.6 \log N_t + \log r_t
\end{equation}
achieves 556,444.1 diggs after 100,000 rounds of simulation, which
is 8.2\% more than $O_1$ and 23.0\% more than $O_3$! However
arbitrary it may seem to give the term $\log N_t$ weight 0.6
rather than 1 is beyond the scope of this paper, but it does show
the complexity of our problem. These experiments demonstrate that
the novelty decay rate needs to be measured with great care, as a
slight change in the decay rate may totally reverse the optimal
order needed to maximize attention.

It is usually hard to analytically compute the performance of a
general index function. For the two simple strategies $O_1$ and
$O_2$, however, some rough estimate can be achieved. For the sake
of generality, assume that there are $m$ positions on the front
page. New stories arrive at a rate $\lambda>0$. Novelty decays as
$r_t = e^{-\alpha t^\beta}$, where $0<\beta\le 1$. Let $\bar a =
\frac 1 m \sum a_i$ be the average position factor, which equals
0.08 for \texttt{digg.com}. Let $\Delta t$ be the refresh time
step, which is 5 minutes for \texttt{digg.com}.

Consider strategy $O_2$ first. According to the index rule, new
stories never appear on the front page. All diggs are generated by
the initial $m$ stories. After time $T$ we have from
Eq.~(\ref{eq:log-growth-rate}) that
\begin{equation}
\log N_T = \sum_{t=0,\Delta t,\dots,T-\Delta t} a_i r_t X_t \Delta t
.
\end{equation}
Hence on average each story's log-performance is
\begin{equation}
E \log N_T = \sum_{t=0, \Delta t, \dots, T-\Delta t} \bar a r_t
\Delta t \approx \bar a \int_0^T r_t dt.
\end{equation}
When $T$ is large, we have
\begin{equation}
\label{eq:log-performance-o1} E \log N_T \approx E\log N_\infty =
\bar a \int_0^\infty r_t dt.
\end{equation}

Next consider $O_1$, which orders the stories by their lifetime. On
average every $s \equiv 1/\lambda$ minutes a new story replaces an
old story, and each old story moves down one position. Hence on
average each story stays on the front page for $ms$ minutes, where
$m$ is the number of positions. We call $ms$ one \emph{page cycle}.
It is the average time it takes to refresh the whole page. We now
see that, before a story disappears from the front page, it
generates
\begin{equation}
N_{ms} = \exp \left( \sum_{t=0,\Delta t,\dots, ms-\Delta t} a_{i(t)}
r_t X_t \Delta t \right)
\end{equation}
diggs, where $i(t)$ is the story's position at time $t$. When an
story gets replaced by a new story, they are counted as one story
restarting from the state $N_t=1$ and $t=0$. The multiplicative
process starts over, and another $N_{ms}$ diggs are generated in
the next $ms$ minutes, on average. Thus, in a total time period
$T$ the process is repeated $T/(ms)$ times, and a total number of
$N_{ms} T/(ms)$ diggs are generated per story. The log-performance
of $O_1$ is approximately
\begin{equation}
\log N_{ms} + \log \left( \frac T{ms} \right) = \sum_{t=0,\Delta
t,\dots, ms-\Delta t} \bar a r_t X_t \Delta t + \log \left( \frac
T{ms} \right),
\end{equation}
where we replaced $a_i(t)$ by $\bar a$ since on average each story
stays in position $1, \dots, m$ for equal times. Taking expectation
on both sides, we have
\begin{equation}
\label{eq:log-performance-o2} E\log N_{ms} + \log \left( \frac T{ms}
\right) \approx \bar a \int_0^{ms} r_t dt + \log \left( \frac T{ms}
\right).
\end{equation}

The critical point can be determined by equating
Eq.~(\ref{eq:log-performance-o1}) and (\ref{eq:log-performance-o2}):
\begin{equation}
\label{eq:critical}E\log N_T - E\log N_{ms} = \log T-\log(ms),
\end{equation}
or
\begin{equation}
\label{eq:critical2}\bar a \int_{ms}^\infty r_t dt = \log \left(
\frac T{ms} \right),
\end{equation}
which holds for any functional form of $r_t$. The left side of
Eq.~(\ref{eq:critical}) can be interpreted as the total novelty
left after a time $ms$, or the total log-performance that can be
gained from one story after one page cycle. The right hand side of
Eq.~(\ref{eq:critical}) is the total log-time left after one page
cycle. Thus, Eq.~(\ref{eq:critical}) and (\ref{eq:critical2}) say
that, after one page cycle, if there is more novelty left than the
log-time remained, the stories should be ordered by decreasing
popularity rather than by decreasing novelty ($O_2$ is better than
$O_1$). Conversely, if novelty decays too fast (not enough novelty
left after one page cycle), then the stories should be ordered by
decreasing novelty rather than decreasing popularity ($O_1$ is
better than $O_2$).

\begin{figure}
\centering
\includegraphics[width=3in]{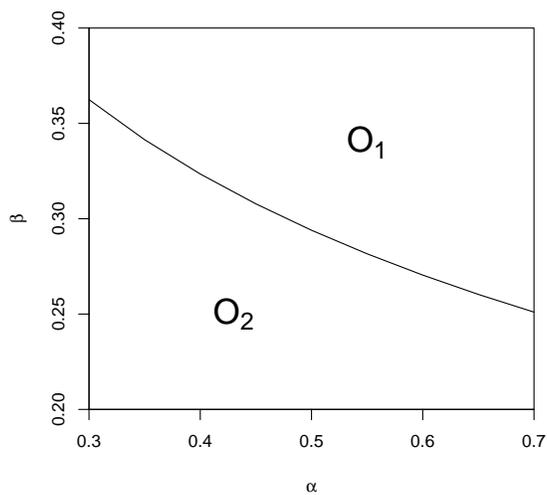}
\caption{\label{fig:phase}The phase diagram. The critical curve is
calculated by solving Eq.~(\ref{eq:critical3}) with $\bar a=0.08$,
$m=15$, $s=20$ and $T=50,000$. When $(\alpha,\beta)$ lies in the
upper half of the phase diagram $O_1$ works better than $O_2$.
Otherwise $O_2$ works better.}
\end{figure}

When $r_t = e^{-\alpha t^\beta}$ it holds that
\begin{equation}
\int_{ms}^\infty r_t dt = \frac{\alpha^{-\frac 1 \beta}}\beta \Gamma
\left( \frac 1 \beta, \,\alpha (ms)^\beta \right),
\end{equation}
where
\begin{equation}
\Gamma(a,x)= \int_x^\infty t^{a-1} e^{-t} dt
\end{equation}
is the \emph{incomplete Gamma function}. In this case the critical
equation can also be written as
\begin{equation}
\label{eq:critical3} \bar a \frac{\alpha^{-\frac 1 \beta}}\beta
\Gamma \left( \frac 1 \beta, \,\alpha (ms)^\beta \right) = \log
\left( \frac T{ms} \right).
\end{equation}
For the parameters of \texttt{digg.com} ($\bar a=0.08$, $m=15$,
$s=20$) and horizon $T=50,000$ one can solve for the critical curve
$(\alpha, \beta)$ on which $O_1$ and $O_2$ have the same
performance. The curve is shown in Fig.~\ref{fig:phase} as a phase
diagram. When the parameters $(\alpha, \beta)$ lie above the
critical curve, the stories should be sorted by $O_1$. Otherwise
they should be sorted by $O_2$.

\begin{figure}
\centering
\includegraphics[width=3in]{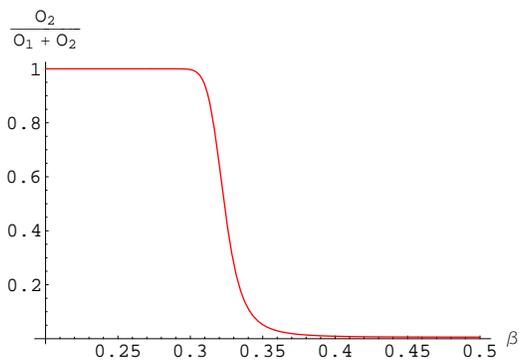}
\caption{\label{fig:phase-transition}The relative performance
$O_2/(O_1+O_2)$ as a function of $\beta$, for fixed $\alpha=.4$.}
\end{figure}

To illustrate how sharp the phase transition is, we plot the
relative performance $O_2/(O_1+O_2)$ as a function of $\beta$, for
fixed $\alpha=.4$, in Fig.~\ref{fig:phase-transition}. As can be
seen, the transition is indeed very sharp.

\section{Conclusion}

In this paper we have shown that depending on the rate of decay of
novelty, two different strategies can be deployed in order to
maximize attention. The first one prioritizes novelty while the
second emphasizes popularity. Most interestingly, the shift from
one to the other as a function of the rate of decay is extremely
sharp, resembling the phase transitions observed in the physical
world.

These results were obtained by focusing on the dynamics of
collective attention and examining the
role that popularity and novelty play in determining the number of
clicks within a given page. In particular, we analyzed three different strategies
that can be deployed in order to maximize attention. The first strategy prioritizes
novelty while the second emphasizes popularity. The third strategy looks
myopically into the future and prioritizes stories that are expected to generate
the most clicks in the next few minutes.  We then showed that the first two
strategies should be selected on the basis of the rate of novelty decay,
while the third strategy performs sub-optimally in most cases. Most interestingly,
we discovered that the relative performance of the first two benchmark strategies
as a function of the rate of novelty decay switches so sharply around
some critical value that it resembles phase transitions observed in
the real world.

Given the importance of maximizing page views for most content
providers, this work suggests a principled way of choosing what to prioritize when designing dynamic websites.
Knowledge of the rates with which novelty and popularity evolve within the website can then be translated into decisions as to what to show first, second, etc.

\end{document}